\documentclass[aps,prl,twocolumn,superscriptaddress,showpacs]{revtex4-1}
\usepackage{amsmath}
\usepackage{graphicx}
\usepackage{url}
\begin{document}
\title{Theoretical Design of a Shallow Donor in Diamond by Lithium-Nitrogen Codoping}

\author{Jonathan E. Moussa}
\email[]{godotalgorithm@gmail.com}
\author{Noa Marom}
\author{Na Sai}
\affiliation{Center for Computational Materials, Institute of Computational Engineering and Sciences, The University of Texas at Austin, Austin, Texas 78712}
\author{James R. Chelikowsky}
\affiliation{Center for Computational Materials, Institute of Computational Engineering and Sciences, The University of Texas at Austin, Austin, Texas 78712}
\affiliation{Departments of Physics and Chemical Engineering, The University of Texas at Austin, Austin, Texas 78712}

\date{\today}

\begin{abstract}
We propose a new substitutional impurity complex in diamond composed of a lithium atom that is tetrahedrally coordinated by four nitrogen atoms (LiN$_4$).
Density functional calculations are consistent with the hydrogenic impurity model,
 both supporting the prediction that this complex is a shallow donor with an activation energy of $0.27 \pm 0.06$ eV.
Three paths to the experimental realization of the LiN$_4$ complex in diamond are proposed and theoretically analyzed.
\end{abstract}

\maketitle

%I. Introduction

%P1 - Why we care about donors in diamond
With respect to multiple figures of merit that estimate
 semiconductor performance in high-power electronics \cite{semiconductor_merit},
 diamond is the best of all known semiconductors.
Steady progress is being made to realize diamond electronics by improving the quality and reducing the cost of single-crystal diamond films
 made by chemical vapor deposition (CVD) \cite{diamond_growth_status}.
Substitutional boron doping has succeeded in producing $p$-type diamond
 and can be incorporated up to concentrations sufficient 
 to cross the metal-insulator transition and produce superconductivity \cite{boron_in_diamond}.
However, no donor impurity has been incorporated into single-crystal diamond
 with sufficiently small activation energy and high concentration to produce an
 $n$-type semiconductor suitable for high-power applications \cite{diamond_dope_review}.

%P2 - Introduce LiN4
In this letter, we propose a new substitutional donor complex in diamond
 composed of lithium tetrahedrally coordinated by nitrogen
 and report on a theoretical study of its activation and formation.
Favorable properties of LiN$_4$ can be inferred from properties of similar structures.
Lithium tetraamine \cite{LiNH3_solid} is a metal composed of Li(NH$_3$)$_4$ molecules
 that are locally isostructural and isoelectronic to LiN$_4$ in diamond.
Each Li(NH$_3$)$_4$ molecule donates an electron to a metallic state permeating the interstitial region between molecules.
If this effect persists for dilute LiN$_4$ in diamond, it should produce shallow donor states with a small activation energy.
A small formation energy is expected for LiN$_4$ based on the high stability of the $B$ center in diamond \cite{N_diamond},
 which is a vacancy ($V$) tetrahedrally coordinated by nitrogen, $V$N$_4$.

\begin{figure}[b]
\includegraphics[width=85mm]{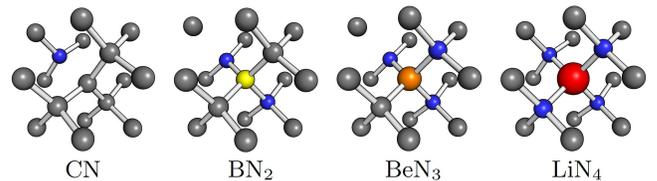}
\caption{\label{defect_fig}(color online) Predicted structures of the $X$N$_n$ donors.
First and second neighbors from $X$ are displayed.
Bonds are omitted for unbonded carbon-nitrogen neighbors, all of which are separated by $2.0$ \AA .
The remaining C-N bonds vary in length from $1.46 - 1.54$ \AA .
The $X$-(C/N) bonds lengthen from right to left on the periodic table:
 $1.50 - 1.60$ \AA \ for B, $1.57 - 1.65$ \AA \ for Be, and $1.72$ \AA \ for Li.}
\end{figure}

%P3 - Connect to past proposals
Our proposal naturally follows from previous codoping proposals \cite{nitrogen_codoping,Si4N_proposal}
 of multi-impurity complexes designed to prevent a carbon-nitrogen bond from breaking near a substitutional nitrogen impurity.
The broken bond forms a deep, localized donor state.
If the bond is preserved, the donor state is predicted to be shallower and more delocalized.
The originally proposed BN$_2$ and newly proposed LiN$_4$ can be connected 
 to nitrogen through a sequence of $X$N$_n$ donor complexes,
 CN $\rightarrow$ BN$_2$ $\rightarrow$ BeN$_3$ $\rightarrow$ LiN$_4$ (Fig. \ref{defect_fig}),
 by reducing the valence of the central atom and electronically compensating with neighboring nitrogens.

%P4 - Reference defects: real and alchemical
The study of other defects provides a useful reference, for comparison to $X$N$_n$ and as a theoretical benchmark.
We consider two well-known substitutional impurities, phosphorus and boron,
 and two artificial defects, C$_5^\mathrm{N}$ and C$_5^\mathrm{B}$.
C$_5^\mathrm{N}$ is a donor formed by adding an electron and compensating the charge
 by replacing a carbon and its four neighbors with fictional nuclei of nuclear charge $Z = 6.2$.
C$_5^\mathrm{B}$ is a similarly constructed acceptor with $Z = 5.8$.
This distribution of nuclear charge preserves the diamond lattice with minimal distortion and enables the formation of shallow defect levels.

%P5 - Shallow impurity model
The minimum activation energy of a point defect is estimated by the hydrogenic impurity model \cite{defect_review}.
In this model, activation is independent of microscopic details of a defect.
It depends only on bulk material properties: the dielectric constant $\epsilon$
 and the ratio between a charge carrier's effective mass $m^*$ \cite{effective_mass_average} and the bare electron mass $m$.
The defect loses a carrier to the nearby band edge,
 where it is Coulombically bound by $13.6 \frac{m^*}{m} \epsilon^{-2}$ eV to the ionized defect.
This produces a donor level below the conduction band edge $E_c$ or an acceptor level above the valence band edge $E_v$
 offset by the binding energy.
Using experimental values \cite{semiconductor_reference}, the model predicts defect levels at
 $E_c - 0.20$ eV and $E_v + 0.45$ eV in diamond compared to $E_c - 0.025$ eV and $E_v + 0.052$ eV in silicon.

%P6 - Success of shallow impurity model
Despite its simplicity, the hydrogenic impurity model is empirically successful.
With acceptor levels measured at $E_v + 0.37$ eV in diamond and $E_v + 0.044$ eV in silicon,
 substitutional boron accurately fits the model.
The substitutional phosphorus donor,
 at $E_c - 0.61$ eV in diamond \cite{P_in_diamond} and $E_c - 0.045$ eV in silicon,
 is considered a shallow donor in silicon but not in diamond.
Shallow donor levels at $E_c - 0.23$ eV have been reported in
 heavily deuterated samples of boron-doped diamond \cite{deuterium_in_B_diamond},
 but with a lifetime too short for applications.

%II. Activation

%P7 - Computational model for impurity levels
A theoretical determination of whether LiN$_4$ is indeed a shallow donor
 requires treatment of microscopic details.
We use a recently proposed method \cite{mixed_defect_method}
 that decomposes the donor activation energy $\Delta_D$
 into a vertical ionization energy and structural relaxation energy,
\begin{align} \label{donor_level}
 \Delta_D  & = E_c + E^+_{\textrm{tot}}(\mathbf{R}^+_D) - E^0_{\textrm{tot}}(\mathbf{R}^0_D) \notag \\
           & = \underbrace{\left[E_c - E_D(\mathbf{R}^+_D) \right]}_{\Delta_D^{\textrm{ionize}}}
             + \underbrace{\left[ E^0_{\textrm{tot}}(\mathbf{R}^+_D) - E^0_{\textrm{tot}}(\mathbf{R}^0_D) \right]}_{\Delta_D^{\textrm{relax}}}.
\end{align}
$E_{\textrm{tot}}^Q(\mathbf{R})$ is the total energy of the system with net charge $Q$ and atomic coordinates $\mathbf{R}$.
$E_D(\mathbf{R})$ is the donor energy level, equal to $E^0_{\textrm{tot}}(\mathbf{R}) - E^+_{\textrm{tot}}(\mathbf{R})$.
The equilibrium atomic coordinates of the donor-containing structure with net charge $Q$ is denoted by $\mathbf{R}^Q_D$.
The corresponding expression for an acceptor activation energy $\Delta_A$ is
\begin{equation} \label{acceptor_level}
 \Delta_A  = \underbrace{\left[E_A(\mathbf{R}^-_A) - E_v \right]}_{\Delta_A^{\textrm{ionize}}}
           + \underbrace{\left[ E^0_{\textrm{tot}}(\mathbf{R}^-_A) - E^0_{\textrm{tot}}(\mathbf{R}^0_A) \right]}_{\Delta_A^{\textrm{relax}}}.
\end{equation}
This decomposition into elementary excitation processes is not unique \cite{na_localize},
 but it enables separate calculations of $\Delta^{\textrm{relax}}$ with total energy methods
 and $\Delta^{\textrm{ionize}}$ with more sophisticated and accurate quasiparticle methods.

%P8 - Condensed basic DFT details
Total energies and equilibrium crystal structures are calculated with density functional theory (DFT) \cite{V,*A,*S,*P,supplement}
 using the Perdew-Burke-Ernzerhof (PBE) functional \cite{PBE}.
C$_5^\mathrm{B}$ and C$_5^\mathrm{N}$ are modeled with alchemical pseudopotentials \cite{PAW_alchemy}.
Isolated defects are approximated with a periodic array of defects in a supercell of diamond.
We use a $6 \times 6 \times 6$ face-centered cubic supercell (432 carbon atoms when defect-free),
 consistent with previous studies \cite{impurity_supercells}.
Supercells with nonzero net charge are simulated with a neutralizing jellium charge distribution.

%P9 - Structural relaxation results
All defect structures are relaxed in their neutral and ionized states from multiple random perturbations of the ideal diamond lattice.
Stable neutral $X$N$_n$ structures are shown in Fig. \ref{defect_fig}.
BN$_2$ and BeN$_3$ also have metastable structures.
Metastable BN$_2$ is similar to its originally predicted structure \cite{nitrogen_codoping},
 with two elongated C-N bonds of length $1.80$ \AA \ rather than a single fully broken C-N bond.
The remaining structures, including metastable BeN$_3$ and all ionized defects,
 produce minor distortions in the diamond lattice that are
 well approximated by one bond length for each bonded pair of elements.

%P10 - Hybrid DFT model for quasiparticle energies
Quasiparticle methods, unlike DFT, are constructed to directly model charge excitation energies such as $\Delta^{\textrm{ionize}}$.
The state-of-the-art is the $GW$ method \cite{GW}, which is too expensive to apply to large supercells at present.
Instead, we use the recently proposed ``PBE-$\epsilon$" method \cite{tuned_hybrid}, which approximates the quasiparticle self-energy as
\begin{align} \label{self_energy}
 \Sigma_{\mathrm{PBE}-\epsilon}(\mathbf{r},\mathbf{r}') = & \left[(1-\epsilon^{-1})v_x^{\mathrm{PBE}}(\mathbf{r}) + v_c^{\mathrm{PBE}}(\mathbf{r}) \right] \delta(\mathbf{r}-\mathbf{r}') \notag \\
                    & - \epsilon^{-1} \rho(\mathbf{r},\mathbf{r}') V(\mathbf{r}-\mathbf{r}'),
\end{align}
 with the PBE exchange $v_x^{\mathrm{PBE}}(\mathbf{r})$ and correlation $v_c^{\mathrm{PBE}}(\mathbf{r})$ potentials 
 and a screened Fock exchange composed of the 1-particle density matrix $\rho(\mathbf{r},\mathbf{r}')$
 and the Coulomb kernel, $V(\mathbf{r}-\mathbf{r}')=e^2/|\mathbf{r}-\mathbf{r}'|$.
With the dielectric constant $\epsilon$ set to the experimental value \cite{semiconductor_reference},
 this method produces a $5.48$ eV band gap for diamond, comparing well to the experimental value of $5.5$ eV.

%P11 - Explain why PBE-epsilon should work
PBE-$\epsilon$ is an adequate quasiparticle method for shallow impurity calculations
 because it approximates the basic physics of a charge carrier bound to an ionized defect.
As in the hydrogenic impurity model, a donor state $\psi_D(\mathbf{r})$ should see an effective Hartree potential
 originating from a screened ionized donor of net charge $\epsilon^{-1}$.
Upon adding a neutral donor to pristine diamond, the Hartree potential is modified by contributions from
 an updated nuclear charge, $\delta \rho_{\textrm{ion}}(\mathbf{r})$, and an added donor electron charge,
\begin{equation}
 \delta v_H(\mathbf{r}) = \int{ V(\mathbf{r}-\mathbf{r}') \left[ |\psi_D(\mathbf{r}')|^2 - \delta \rho_{\textrm{ion}}(\mathbf{r}') \right] d\mathbf{r}'}.
\end{equation}
The donor state also sees a modified effective potential from its self interaction in the screened Fock exchange,
\begin{align}
 -\int & \epsilon^{-1} \psi_D(\mathbf{r})\psi_D^*(\mathbf{r}') V(\mathbf{r}-\mathbf{r}') \psi_D(\mathbf{r}') d\mathbf{r}' = \delta v_{sX}(\mathbf{r}) \psi_D(\mathbf{r}) \notag \\
  &\delta v_{sX}(\mathbf{r}) = - \epsilon^{-1} \int{ V(\mathbf{r}-\mathbf{r}') |\psi_D(\mathbf{r}')|^2 d\mathbf{r}'}.
\end{align}
The total donor-induced potential is $\delta v_{H}(\mathbf{r}) + \delta v_{sX}(\mathbf{r})$,
 which corresponds to the bare ionized donor $\delta \rho_{\textrm{ion}}(\mathbf{r})$
 and an effective screening cloud $(\epsilon^{-1}-1)|\psi_D(\mathbf{r})|^2$.
A similar argument applies to acceptors.
The screening cloud has the right net charge but the wrong length scale: the Bohr radius of the donor state is $6.3$ \AA,
  but the screening length in diamond is estimated to be $1.5$ \AA \cite{model_dielectric}.
We find PBE-$\epsilon$ to be a good compromise between costly $GW$ corrections to the screening length and PBE without Fock exchange,
 which has an effective screening cloud of unit charge
 that suppresses the long-range electron-impurity interaction and 
 produces a donor impurity band nearly degenerate with the conduction band edge \cite{supplement}.

%P12 - Finite-size effects
The periodic array of defects broadens defect levels into bands up to $0.4$ eV in width.
Modeling or extrapolation is necessary to extract an accurate activation energy.
We use a tight-binding ansatz and a range of supercells from $5 \times 5 \times 5$ to $8 \times 8 \times 8$
 for extrapolation, which is described in detail in the Supplemental Material \cite{supplement}.

\begin{table}
\caption{\label{impurity_table} Donor and acceptor activation energies calculated with both the marker method \cite{impurity_supercells} and PBE-$\epsilon$ quasiparticles, compared to experiment.
PBE-$\epsilon$ results are separated into relaxation and ionization contributions,
 $\Delta^{\mathrm{PBE}-\epsilon} = \Delta^{\mathrm{ionize}} + \Delta^{\mathrm{relax}}$,
 as in Eqs. (\ref{donor_level}) and (\ref{acceptor_level}).
$\delta \Delta^{\mathrm{ionize}}$ is the RMS variance of extrapolation \cite{supplement}.
All energies are in units of eV.}
\begin{ruledtabular}
\begin{tabular}{c | c c c | c c c }
 Defect & $\Delta^{\mathrm{exp}}$ & $\Delta^{\mathrm{marker}}$ & $\Delta^{\mathrm{PBE}-\epsilon}$ & $\Delta^{\mathrm{ionize}}$ & $\delta \Delta^{\mathrm{ionize}}$ & $\Delta^{\mathrm{relax}}$ \\  \hline
 C$_5^\mathrm{N}$ & $\cdots$   & 0.45     & 0.31 & 0.31 & 0.03 & 0.00 \\
 LiN$_4$          & $\cdots$  & 0.48     & 0.27 & 0.27 & 0.03 & 0.00 \\
 BeN$_3$\footnotemark[1]        & $\cdots$    & 0.56     & 0.40 & 0.39 & 0.04 & 0.01 \\
 P                & 0.61 & \ \ 0.61\footnotemark[2] & 0.56 & 0.54 & 0.02 & 0.02 \\
 BeN$_3$          & $\cdots$    & 0.78     & 0.62 & 0.39 & 0.04 & 0.23 \\
 BN$_2$\footnotemark[1]         & $\cdots$    & 0.88     & 0.77 & 0.50 & 0.03 & 0.27 \\
 BN$_2$           & $\cdots$   & 1.30     & 1.19 & 0.50 & 0.03 & 0.69 \\
 N                & 1.7  & 1.67     & 1.71 & 0.86 & 0.04 & 0.85 \\ \hline
 C$_5^\mathrm{B}$ & $\cdots$    & 0.31     & 0.30 & 0.30 & 0.01 & 0.00 \\
 B                & 0.37 & \ \ 0.37\footnotemark[2] & 0.31 & 0.31 & 0.03 & 0.00
\footnotetext[1]{Metastable structure.}
\footnotetext[2]{Experimental marker.}
\end{tabular}
\end{ruledtabular}
\end{table}

%P13 - Discuss results, inherent errors, and strength of conclusions.
Theoretical activation energies are listed in Table \ref{impurity_table}
 alongside known experimental values.
The PBE-$\epsilon$ quasiparticle approach is compared
 to the semi-empirical marker method \cite{impurity_supercells},
 which calculates activation energies relative to an experimental ``marker'' impurity
 using PBE total energy differences.
The marker method predicts larger activation energies than PBE-$\epsilon$.
These deviations grow with decreasing activation energy
 and become as large as the value we are attempting to predict.
This can be explained by delocalization errors in PBE
 that are reduced in PBE-$\epsilon$ with the addition of Fock exchange \cite{na_localize}.
Therefore, PBE-$\epsilon$ should be more reliable than the marker method as a predictor of activation energies over a wider energy range.
Doubling the extrapolation variance provides a wide enough confidence interval
 for the PBE-$\epsilon$ predictions to be consistent with all experiments.
LiN$_4$ is shallower than the artificial shallow donor C$_5^\textrm{N}$
 and an activation energy of $0.27 \pm 0.06$ eV is consistent with the hydrogenic impurity model.
We conclude that LiN$_4$ is a shallow donor.

%III. Formation

%P14 - Summarize possible methods to incorporate LiN4 into diamond
Having confirmed the viability of LiN$_4$ as a shallow donor in diamond, we now consider three synthesis paths.
The first path is the diffusion of lithium into diamond \cite{Li_implanted} with a high concentration of $B$ centers ($V$N$_4$).
The second path is high-pressure, high-temperature (HPHT) diamond synthesis \cite{HPHT_from_organic}
 in the presence of lithium and nitrogen.
The third path is CVD diamond synthesis with the LiN$_4$ impurity preformed in
 a seed material \cite{CVD_seeds} or deposited molecule \cite{Li_in_CVD}.

%P15 - Li diffusion scenarios: competing trapping sites, none of which are strongly favored.
Nitrogen incorporates substitutionally into diamond, found as an isolated center, a dimer, or clustered around a vacancy, $V$N$_m$ \cite{N_diamond}.
High temperature treatment causes vacancies to become mobile and
  cluster with nitrogen to form mobile $V$N$_m$ complexes.
Theoretical studies of Li diffusion in N-free diamond \cite{Li_theory1}
 predict the interstitial (Li$_i$) to be a mobile donor that is strongly trapped by vacancies.
The natural extension to N-rich diamond is a general trapping process, Li$_i$ + $V$N$_m$ $\rightarrow$ LiN$_m$,
 which we calculate to bind at $6.88$, $7.24$, $8.04$, $8.37$, and $6.08$ eV for $m = 0,\ldots,4$.
All sites trap strongly, but $V$N$_4$ is preferred least by Li$_i$.
The LiN$_m$ defect sequence has a regular trend of activity from triple acceptor ($m=0$) to single donor ($m=4$).
Assuming all vacancies will be filled with lithium and the only acceptors are Li, LiN, and LiN$_2$,
 then the defect concentrations $n(X)$ must satisfy the inequality
\begin{equation} n(\textrm{LiN}_4) > 3 n(\textrm{Li}) + 2 n(\textrm{LiN}) + n(\textrm{LiN}_2) \end{equation}
 to prevent all LiN$_4$ from being passivated.
Therefore, lithium diffusion into a diamond sample is only likely to succeed in producing active LiN$_4$
 if the average number of nitrogens around each vacancy in the pre-lithiated sample is greater than 3.

%P16 - LiN4 in HPHT: an exercise in failure.
HPHT synthesis of LiN$_4$ at a detectable concentration
 requires sufficient thermodynamic stability of the complex at an accessible pressure and temperature.
At zero temperature, we have found two pairwise decomposition processes that passivate shallow donor activity,
\begin{subequations}\begin{align}
     2 \textrm{LiN}_4 &\rightarrow (\textrm{LiN}_4)_2 \\
 2 \textrm{LiN}_4 + V &\rightarrow \textrm{LiN}_3 + \textrm{Li}V\textrm{N}_5. \label{badguy}
\end{align}\end{subequations}
The first reaction produces a LiN$_4$ dimer with neighboring nitrogens that break the N-N bond,
 which only lowers enthalpy below 530 GPa.
The second reaction exchanges
 a nitrogen and binds an additional vacancy to the N-rich complex,
 which produces an octahedrally coordinated Li surrounded by CN$_5$.
Assuming a zero chemical potential for $V$, this process lowers enthalpy at all tested pressures (up to 700 GPa)
 and has a minimum enthalpy reduction of 2.47 eV at 210 GPa.
As a result of the process in Eq. (\ref{badguy}),
 it is unlikely that LiN$_4$ can be synthesized in HPHT or any other conditions
 that enable LiN$_4$ and $V$ to become mobile and interact with each other.

\begin{figure}
\includegraphics[width=86mm]{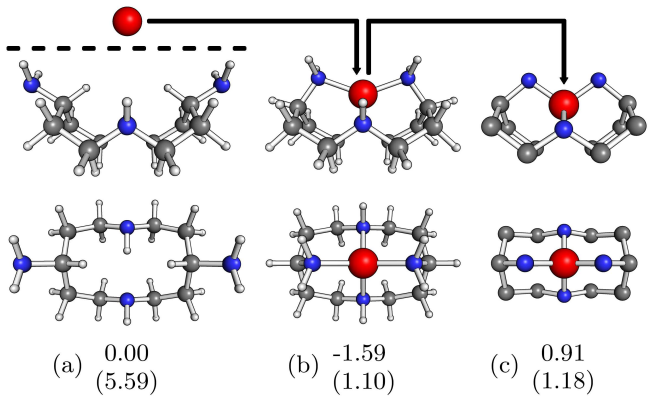}
\caption{\label{molecule_fig} (color online) Top and side views of (a) isolated Li and 1,7-diazacyclododecane-4,10-diamine,
 (b) Li bound to 1,7-diazacyclododecane-4,10-diamine, and (c) Li bound to the $V$N$_4$ defect in diamond.
Relative formation energies of Li (and Li$^+$ in parentheses) are reported in eV,
 from PBE total energies (and PBE-$\epsilon$ ionization energy for Li$^+$ in (c)).
The structures for Li$^+$ are similar to the neutral structures shown.}
\end{figure}

%P17 - Molecular precursor design.
Formation of the LiN$_4$ complex in a CVD process from separate lithium and nitrogen sources is likely to be a rare event
 because it involves a coincidence of five atoms, each with a presumably low concentration.
This problem can be avoided by preforming the complex within a precursor molecule.
A suitable LiN$_4$ precursor should be small to enhance volatility and simplify synthesis,
 closely conform to the diamond lattice it is to be incorporated into,
 and exist as a well-defined lithium-free molecule that strongly binds a lithium atom or ion.
Diamondoids \cite{diamondoids} satisfy the second constraint
 and many chelants \cite{cryptand} satisfy the third constraint,
 but we propose a new analog of cyclododecane (Fig. \ref{molecule_fig}) that satisfies all three constraints
 (with IUPAC name 1,7-diazacyclododecane-4,10-diamine).
Lithiation of this molecule should abstract H (as $\frac{1}{2}$H$_2$)
 to form a more stable (by $0.24$ eV in our calculations) lithamide.

%P18 - LiN4 in basic CVD scenarios: a very incomplete picture of metastable Li positions.
Figure \ref{molecule_fig} depicts lithium in three metastable positions relevant to CVD: the isolated atom,
 bound to a precursor molecule, and bound to a $B$ center in diamond.
Li$^+$ is strongly bound to both sites.
Li is not bound to the $B$ center,
 but will remain trapped there because of a large energy barrier.
The Li-N bond length of $2.0$ \AA \ within the precursor is reduced to $1.72$ \AA \ within the $B$ center.
Bond strain can be quantified indirectly by comparing total energies of the relaxed $V$N$_4$ cavity
 to the LiN$_4$ defect structure with Li removed, which results in a difference of $0.27$ eV.
The reduction of LiN$_4$'s ionization energy from the precursor to diamond
 is caused by a destabilization of the neutral state rather than any significant change of the ionized state.

%P19 - CVD caveat: does the internal structure of the precursor survive? [3 LINES]
Successful CVD synthesis of LiN$_4$ is contingent on the existence of growth conditions
 that preserve the internal structure of the LiN$_4$ precursor while still enabling good diamond crystal formation,
 which is an open problem.

%P20 - Brief conclusions
In short, we propose LiN$_4$ as a new donor complex in diamond
 with a predicted activation energy of $0.27 \pm 0.06$ eV.
Synthesis of LiN$_4$ is likely to require that Li be reacted with a preformed $V$N$_4$ complex,
 either within diamond or a precursor molecule.
While further studies of Li-N-$V$ defect chemistry in diamond are warranted,
 the present result should serve as sufficient impetus for the pursuit of experimental realization of LiN$_4$.

\begin{acknowledgments}
We acknowledge support from the National Science Foundation under Grant No. DMR-0941645.
Computing resources were provided by the National Energy Research Scientific Computing Center (NERSC).
Alchemical pseudopotentials and \textsc{vasp} modifications required for their use were kindly provided by Daniel Sheppard.
Molecular visualizations were generated with \textsc{pymol} \cite{pymol}.
J.E.M. thanks Jay Deep Sau for helpful discussions.
\end{acknowledgments}

%
%\end{document}

\newpage

%\title{Supplemental Material to: Theoretical Design of a Shallow Donor in Diamond by Lithium-Nitrogen Codoping}

%\author{Jonathan E. Moussa}
%\email[]{godotalgorithm@gmail.com}
%\author{Noa Marom}
%\author{Na Sai}
%\affiliation{Center for Computational Materials, Institute of Computational Engineering and Sciences, The University of Texas at Austin, Austin, Texas 78712}
%\author{James R. Chelikowsky}
%\affiliation{Center for Computational Materials, Institute of Computational Engineering and Sciences, The University of Texas at Austin, Austin, Texas 78712}
%\affiliation{Departments of Physics and Chemical Engineering, The University of Texas at Austin, Austin, Texas 78712}

%\date{\today}

%\maketitle

\textbf{Supplemental Material to ``Theoretical Design of a Shallow Donor in Diamond by Lithium-Nitrogen Codoping''}

%P1 - Further details for ground state studies
All DFT calculations are performed using version 5.2 of the Vienna Ab-initio Simulation Package (\textsc{vasp}) \cite{V2,*A2,*S2,*P2}.
A planewave cutoff of $400$ eV is used for all calculations,
 along with the manual-recommended \cite{VASP_manual2} projector augmented-wave pseudopotentials
 (or alchemical mixtures thereof \cite{PAW_alchemy2}).
Band occupations are set by Gaussian smearing with a width of $0.01$ eV.
Relaxed structures have the forces on all atoms reduced below 0.1 eV/\AA.
The PBE lattice constant of 3.57 \AA \ is used for all calculations in diamond at zero pressure.
Finite pressure is simulated by reducing the lattice constant.
A $2 \times 2 \times 2$ Monkhorst-Pack grid is used to sample the Brillouin zone (BZ) of diamond supercells for PBE total energy calculations,
 as suggested in previous studies \cite{impurity_supercells2}.
All molecules are simulated at the $\Gamma$-point of a cubic supercell with a $20$ \AA \ lattice constant.
Predictions of molecular structure are made by generating many locally stable conformations
 using the MMFF94s force field in \textsc{avogadro} \cite{avogadro2} and further relaxing them in \textsc{vasp}.
The lowest energy structure found by this search is presumed to be the ground state.

%P2 - Fock exchange calculation details
The computational cost of PBE-$\epsilon$ calculations is
 larger than PBE because of the Fock exchange step.
The ratio between costs grows linearly with the number of points used to sample the BZ,
 which limits our PBE-$\epsilon$ calculations to a single BZ point at a time for large supercells.
When computing the band structure at a BZ point, the density matrix is constructed
 from only that BZ point and iterated to self-consistency.
This is a systematic error for finite supercells, but it is exponentially suppressed
 with increasing supercell size because all bands are fully occupied
 (half-filled impurity bands are fully occupied in one spin channel and empty in the other).

\begin{figure}
\includegraphics[width=86mm]{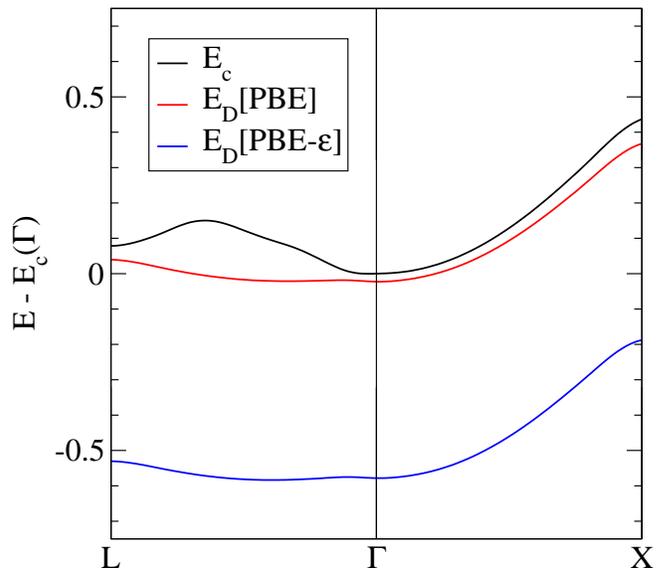}
\caption{\label{hybrid_band_fig} PBE and PBE-$\epsilon$ band structures of LiN$_4$ in a $6 \times 6 \times 6$ supercell
 in the spin channel where the donor impurity band is occupied.
Only the donor impurity band and the lowest branch of the conduction band are shown.
The conduction band is similar at both levels of theory and only one is plotted.}
\end{figure}

%P3 - Examination of bands
The addition of Fock exchange in switching from PBE to PBE-$\epsilon$
 has the basic effect of making the impurity bands deeper and more sensitive to finite-size effects.
This is illustrated in Fig. \ref{hybrid_band_fig} with a plot of the lowest branch of the conduction band
 and the donor impurity band of LiN$_4$ in a $6 \times 6 \times 6$ supercell along the $L-\Gamma-X$ high symmetry path.
PBE and PBE-$\epsilon$ produce similar results, except that the PBE-$\epsilon$ impurity band
 is rigidly shifted downwards in energy by $0.68$ eV.
The average depth of the PBE impurity band is smaller than the expected shallow impurity depth of $0.2$ eV
 and does not change significantly in larger supercells.
The PBE-$\epsilon$ impurity band is deeper than expected,
 but it becomes more shallow with increasing supercell size and decreasing impurity density.
To estimate the isolated impurity limit, some kind of extrapolation must be performed.

%F1 - Band plot: PBE vs. PBE-epsilon for donor band vs. conduction band of LiN4 in 6^3 supercell

%P4 - Activation energy calculations and extrapolation
We calculate impurity ionization energies $\Delta^{\textrm{ionize}}$ by using
 a range of diamond supercells from $5 \times 5 \times 5$ to $8 \times 8 \times 8$.
Smaller supercells have finite-size effects that are too large and complicated
 to be fit to a simple extrapolation model and larger supercells are computationally intractable.
We assume a tight binding picture where impurity states are localized about each impurity
 and an impurity band manifests from hopping between impurity states.
The hopping energy and impurity bandwidth should decrease exponentially
 with increasing distance between neighboring impurities $R$.
All  finite-size effects in the impurity band should follow a simple exponential form $\propto \exp(-R/R_0)$.
$R_0$ is an estimate of the effective Bohr radius of the impurity state,
 but it may be partially contaminated by finite-size BZ sampling errors.
If the finite-size dependence is similar and simple enough,
 then some combination of BZ points in the impurity band should cancel the finite-size effects.
Specifically, we make a simple but arbitrary choice to define
\begin{subequations}\label{extrap}\begin{align}
 \Delta_D^{\textrm{ionize}} &\approx E_c(\Gamma) - \left[ (1-x) E_D(\Gamma) + x E_D(X) \right]\\
 \Delta_A^{\textrm{ionize}} &\approx \left[ (1-x) E_A(\Gamma) + x E_A(X) \right] - E_v(\Gamma),
\end{align}\end{subequations}
 where $E_c$ is the lowest branch of the conduction band, $E_v$ is the highest branch of the valence band,
 $E_D$ is the occupied donor impurity band, and $E_A$ is the unoccupied acceptor impurity band.
For large supercells, this expression will be independent of the choice of $x$.
With this free parameter and the results from 4 supercells,
 we perform a least squares fit to minimize the size-dependence of $\Delta^{\textrm{ionize}}$
 in Eq. (\ref{extrap}) about an asymptotic ionization energy.
Additionally, we fit $E_D(X) - E_D(\Gamma)$ and $E_A(\Gamma) - E_A(X)$ to the form $C \exp(-R/R_0)$ to estimate $R_0$.

%P5 - Discussion of results
Extrapolation results are tabulated in Table \ref{extrap_table}.
$\delta \Delta^{\mathrm{ionize}}$ is the root-mean-square (RMS) deviation
 between the extrapolated $\Delta^{\textrm{ionize}}$ and
 Eq. (\ref{extrap}) for the 4 calculated supercells.
The extrapolation procedure succeeds in reducing the variance to a value significantly below
 the bandwidth of the impurity bands in the calculated supercells.
The effective Bohr radii extracted from extrapolation increase with decreasing ionization energy,
 as expected from a simple hydrogenic model of impurities.
The hydrogenic impurity model gives a radius of $6.3$ \AA \ for shallow donors and $2.8$ \AA \ for shallow acceptors,
 which is smaller than the extrapolated values and especially so for the acceptors.
The cause of this deviation is unclear, but the effect on the ionization energies seems minor
 since boron is within $0.06$ eV of its experimental activation energy of $0.37$ eV.

%T1 - Extrapolation table
%R_0 , ionization energy , rms error in ionization energy.
\begin{table}
\caption{\label{extrap_table} Extrapolated donor and acceptor properties. $R_0$ is the effective Bohr radius in \AA.
$\Delta^{\mathrm{ionize}}$ is the ionization energy and $\delta \Delta^{\mathrm{ionize}}$ is its RMS variance, both in eV.
$x$ is the parameter used in the extrapolation formulae in Eq. (\ref{extrap}).}
\begin{ruledtabular}
\begin{tabular}{c | c c c c}
 Defect           & $R_0$ & $\Delta^{\mathrm{ionize}}$ & $\delta \Delta^{\mathrm{ionize}}$ & $x$ \\  \hline
 C$_5^\mathrm{N}$ & 7.9   & 0.31 & 0.03 & 0.74 \\
 LiN$_4$          & 7.8   & 0.27 & 0.03 & 0.74 \\
 BeN$_3$          & 6.7   & 0.39 & 0.04 & 0.69 \\
 BN$_2$           & 5.5   & 0.50 & 0.03 & 0.62 \\
 P                & 5.2   & 0.54 & 0.02 & 0.61 \\
 N                & 4.5   & 0.86 & 0.04 & 1.04 \\ \hline
 B                & 7.9   & 0.31 & 0.03 & 1.05 \\
 C$_5^\mathrm{B}$ & 7.0   & 0.30 & 0.01 & 0.95 \\
\end{tabular}
\end{ruledtabular}
\end{table}


\begin{thebibliography}{34}
\bibitem [{eff()}]{effective_mass_average}%
  \BibitemOpen
  \href@noop {} {}\bibinfo {note} {For electrons, $m_e^*$ is averaged assuming
  an anisotropic conduction band as $(m_e^*)^{-1} = [(m_{e\parallel}^*)^{-1} +
  2 (m_{e\perp}^*)^{-1}]/3$. For holes, $m_h^*$ is taken from the heavy hole
  band.}\BibitemShut {Stop}%
\end{thebibliography}
\end{document}